\documentclass[fleqn,10pt]{wlscirep}
\usepackage[utf8]{inputenc}
\usepackage[T1]{fontenc}
\usepackage{chemformula}
\usepackage{caption}
\usepackage{subcaption}
\usepackage{wrapfig}
\usepackage{gensymb}
\usepackage{hyperref}
\newcommand{\bvec}[1]{\boldsymbol{#1}}

\title{Surface Energy Driven Grain Growth Model: FePt L1$_0$ Nanoparticles}

\author[1,*]{Connor Skelland}
\author[1]{Gino Hrkac}
\affil[1]{College of Engineering, Mathematics, and Physical Sciences, University of Exeter, Exeter, Devon, United Kingdom}
\affil[*]{connor.publications@proton.me}

\begin{abstract}
We developed a grain growth model that is based on the energy minimisation of surfaces with respect to the volume energy and the grain's environment. We used the well known FePt L1$_\text{0}$ system to discover the physical factors that drive the shape and size of FePt grains. It was found that the preferred growth directions are along symmetry planes that are determined by the basic crystal and driven by surface energy minimisation. The model developed here can be used to predict a grain's growth and shape as a function of atomic number and composition. This means that by tailoring a grain's surface and grain boundaries the shape of the magnetic grains can be manipulated.

\end{abstract}
\begin{document}

\flushbottom
\maketitle

\section*{Introduction}

Magnetic materials have a long history of driving technological growth and societal progress\cite{verschuur1996hidden}. In the modern world, three of the most important industries, Green-Energy, the Automotive-Sector, and Magnetic Recording are reliant on the improvement of at least one of a magnetic materials fundamental properties - coercivity, anisotropy, stability, etc. Throughout history, the greatest leaps in improvement have come from step changes in performance caused by the discovery of impressive new crystal structures. For example, the transition from Steel to Alnico, and Alnico to Rare-Earth based magnetic materials\cite{livingston1990history}.

However, since the discovery of \ch{Nd2Fe14B} in 1984\cite{sagawa1984new}, the hard permanent magnet research community has struggled to find new crystal structures with improved magnetic properties and a compelling price point. In the intervening time, the focus has shifted to optimising material properties via microstructure engineering, which is achieved through alterations to fabrication process and minor alterations to chemical composition\cite{varaprasad2013microstructure, kronmuller1996micromagnetism, goll1998magnetic, woodcock2012understanding, exl2018magnetic, ener2021twins}. The success of this method is exemplified by the anistropic sintered \ch{Nd2Fe14B} magnets, detailed in a patent by Sakuma et al.\cite{sakuma2016}.

The success of microstructure engineering demonstrates the importance of controlling magnetism through structural control. The implication of this is that physical models must capture defects, phase inhomogeneities, grain shape, grain size, and grain orientation if they are to reproduce realistic magnetic effects.

Currently microstructure informed modelling is limited - a combination of Voronoi tesselation and integranular phase definition is the most accurate template used for both micromagnetic and spin dynamic modelling \cite{fischbacher2018micromagnetics, evans2014atomistic, caballero2018magnetic, sankaran2018evidence, zheng2003atomistic}. Whilst capable of capturing the general size and rough shape of grains, these models are limited by common simplifications such as: intergranular phases are indistinguishable in their magnetic properties, all intergranular phases are the same width, and multi-phase grains share average values in anisotropy and exchange. Although reasonable, these simplifications prevent the models from totally capturing grain boundary effects that cause domain wall pinning, reverse grain nucleation, and intergranular exchange. These effects have a prominent role in two of the most important magnetic crystal structures: \ch{Nd2Fe14B}\cite{hrkac2010role, woodcock2012understanding} and FePt L1$_0$\cite{sepehri2017pt} and therefore must be taken into consideration. To do this, models must accurately represent microstructure.

To address this problem, we took the view that accurate representation of microstructure starts with an understanding of what drives its development. This led us to produce a grain growth model that let us investigate how surface energy defines preferential grain shapes, and how preferential shapes change with grain size. We applied this model to FePt L1$_0$ for two reasons, firstly, it is a well studied structure, and secondly, grain size and shape are crucial to its application in the magnetic hard drive industry\cite{varvaro2014l10, mosendz2012ultra}. Particularly, grains homogeneous in size and shape improve the signal to noise ratio of hard drive recording tracks\cite{piramanayagam2007perpendicular, miles2007effect}, decreasing the number of grains required per bit and increasing information density as a result. 

Our model uses Molecular-Dynamics-based energy minimisation to calculate the preferability of grains composed of identical elements; their identical composition makes the grains different states of the same thermodynamic system. Using the Boltzmann distribution of the underlying system, we assigned an energy based probability to each of the grains (which are considered states within the same Boltzmann distribution), which made it possible to establish the relative probability of each type of grain - see the methodology section. This is a first step towards a general methodology for producing realistic atomistic models, that can be used as input to Heisenberg spin models and finite element based micromagnetic modelling. Future applications of this model will include the investigation of granular rare-earth intermetallics for the automotive industry.

\section*{Results}

\begin{figure}[!t]
    \captionsetup[subfigure]{font=Large,justification=justified,singlelinecheck=false,labelfont={rm, bf},labelformat=simple,}
    \newcommand{\vspaced}{\vspace{-0.8em}}
    \newcommand{\vvspaced}{\vspace{-1.6em}}
    \centering
    \begin{subfigure}[t]{0.4\textwidth}
        \centering
        \subcaption{}
        \vspaced
        \includegraphics[width=\textwidth]{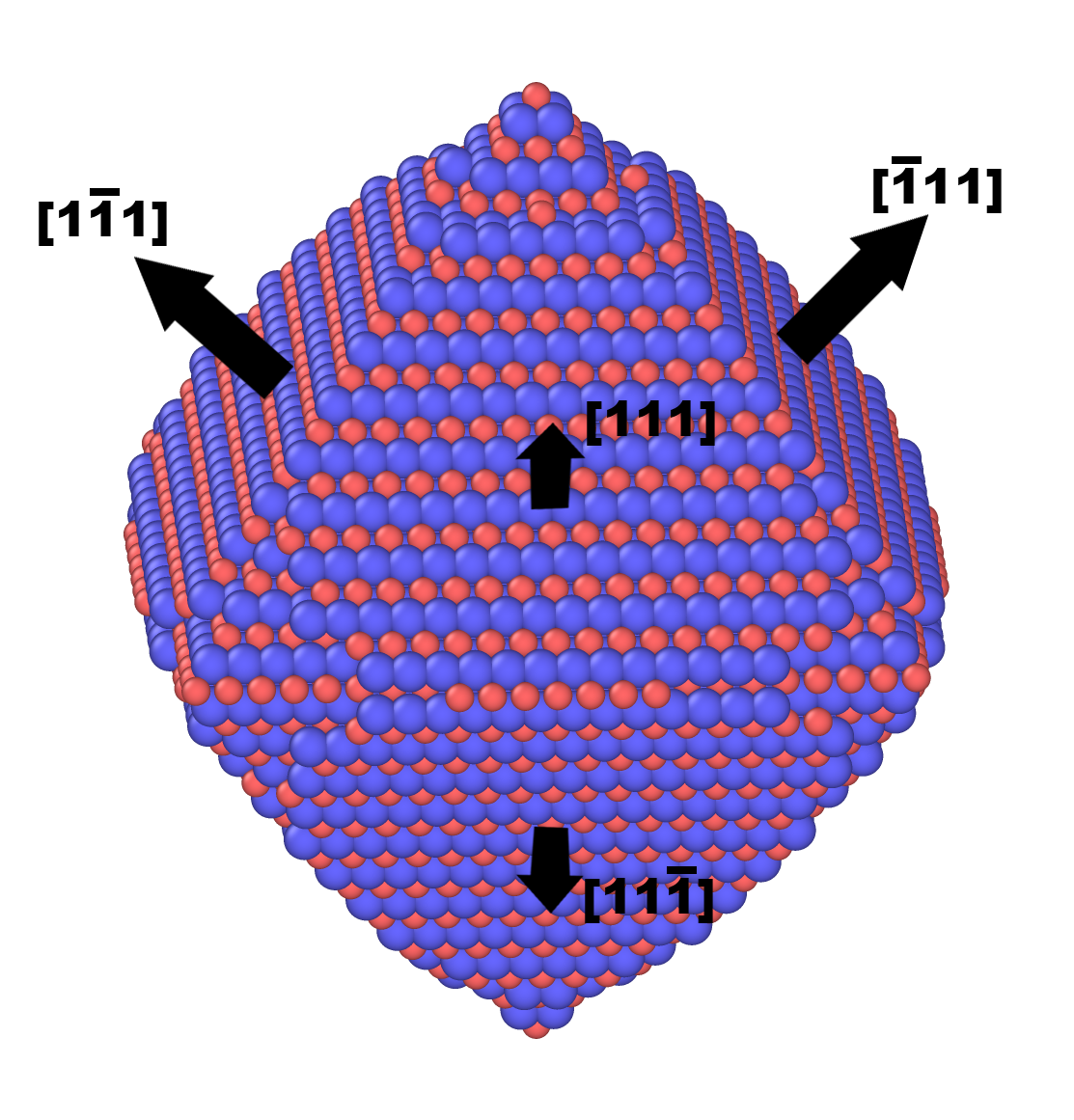}
        \label{fig:fept_l10_octahedron_10000_atoms_directions}
        \vvspaced
    \end{subfigure}
    \hspace{0.18\textwidth}
    \begin{subfigure}[t]{0.4\textwidth}
        \centering
        \subcaption{}
        \vspaced
        \includegraphics[width=\textwidth]{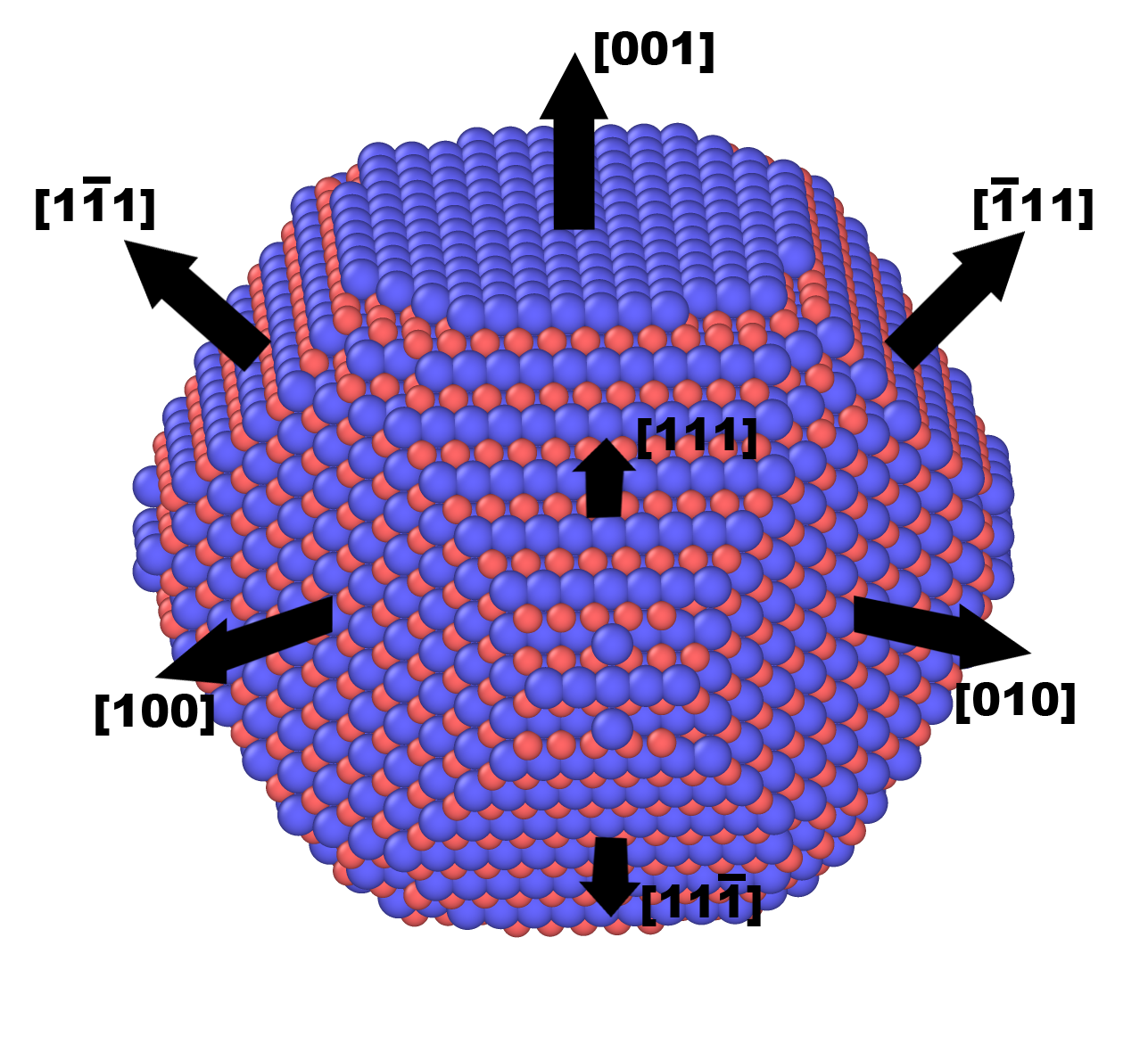}
        \label{fig:fept_l10_truncated_octahedron_major_10000_atoms_directions}
    \end{subfigure}
    \begin{subfigure}[t]{0.4\textwidth}
        \centering
        \subcaption{}
        \vspaced
        \includegraphics[width=\textwidth]{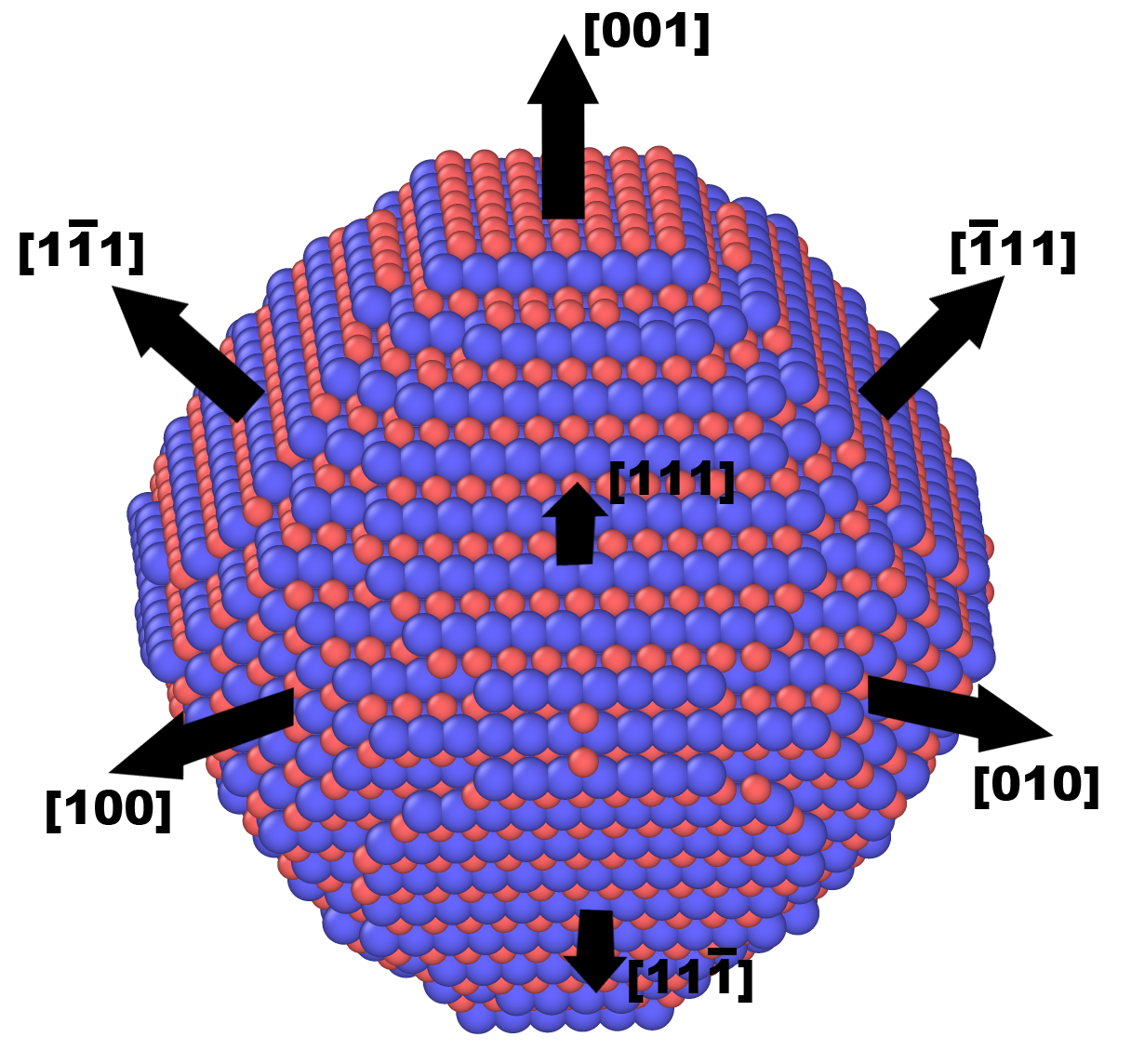}
        \label{fig:fept_l10_truncated_octahedron_minor_10000_atoms_directions}
    \end{subfigure}
    \caption{Compositionally matched grain morhpologies at the size of 10,000 atoms. Showing a) Octahedron, b) Truncated Octahedron Major, and c) Truncated Octahedron Minor. The arrows represent the grain's surface normals.}
    \label{fig:fept_l10_surface_energy_informed_grains}
\end{figure}

To produce realistic grain shapes, we calculated the surface energies of the five lowest index symmetry planes of FePt L1$_0$ in LAMMPS, the results of this can be seen in Table \ref{tab:fept_l10_surface_energies} along with a comparison to calculations by Kim et al.\cite{kim2006origin}. We used these surface energies to generate five types of grain shape across a range of sizes.
\begin{table}[!h]
    \def\arraystretch{1.1}
    \newcommand{\vspaced}{\vspace{-0.8em}}
    \newcommand{\vvspaced}{\vspace{-1.6em}}
    \centering
    \vspace{0.5em}
    \begin{tabular}{|c|c|c|}
        \hline
        Plane & Our Surface Energy (erg/cm$^2$) & Kim's Surface Energy (erg/cm$^2$)\\
        \hline
        111 & 2200 & 2198\\
        101 & 2685 & 2714\\ 
        110 & 2756 & 2650\\
        100 & 2449 & 2719\\
        001 & 2919 & 2740\\
        \hline
    \end{tabular}
    \caption{Surface Energies of low index FePt L1$_\text{0}$ planes, taken from Kim et al.\cite{kim2006origin}, compared to our own calculated values.}
    \vvspaced
    \vspace{0.5em}
    \label{tab:fept_l10_surface_energies}
\end{table}

\begin{figure}
    \captionsetup[subfigure]{font=Large,justification=justified,singlelinecheck=false,labelfont={rm, bf},labelformat=simple,}
    \newcommand{\vspaced}{\vspace{-0.8em}}
    \newcommand{\vvspaced}{\vspace{-1.6em}}
    \centering
    \begin{subfigure}[t]{0.4\textwidth}
        \centering
        \subcaption{}
        \vspaced
        \includegraphics[width=\textwidth]{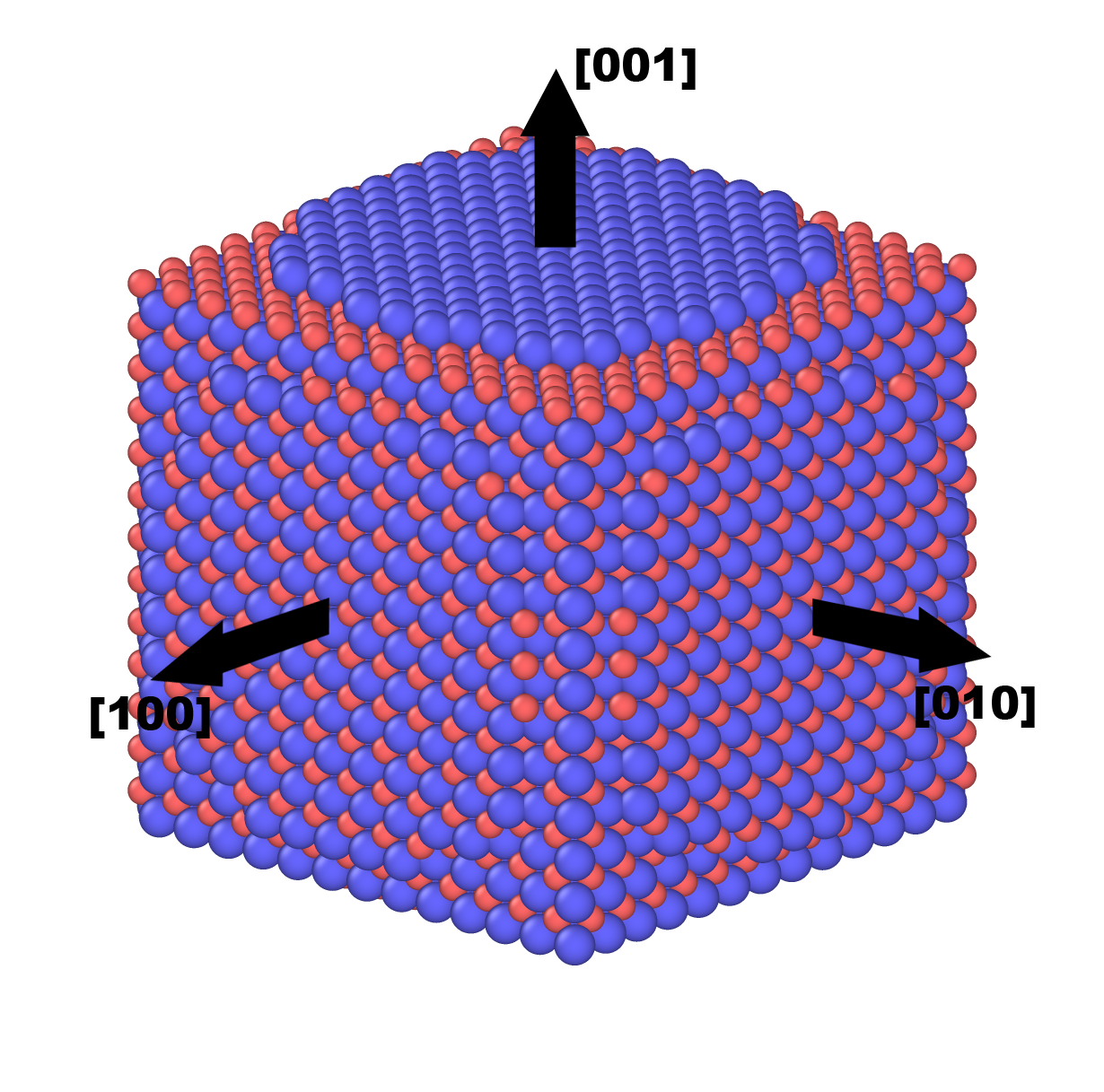}
        \label{fig:fept_l10_cuboid_10000_atoms_directions}
    \end{subfigure}
    \begin{subfigure}[t]{0.35\textwidth}
        \centering
        \subcaption{}
        \vspaced
        \includegraphics[width=\textwidth]{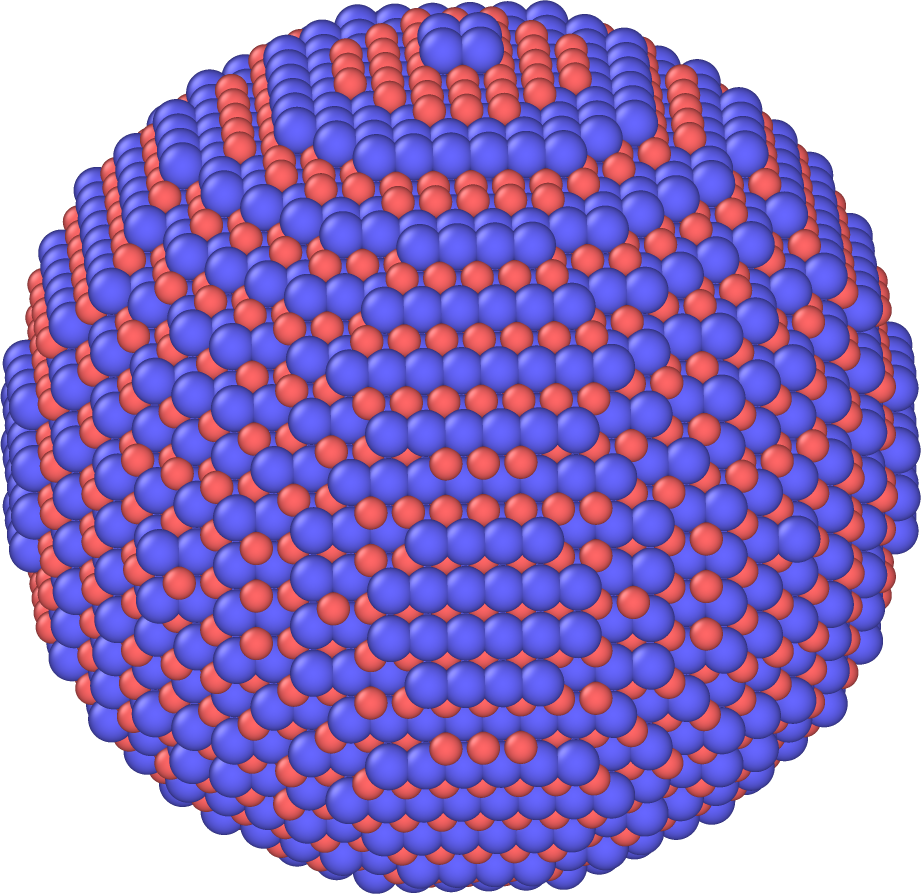}
        \label{fig:fept_l10_sphere_10000_atoms_directions}
    \end{subfigure}
    \caption{Compositionally matched grain morhpologies at the size of 10,000 atoms. Showing a) Cuboid, and b) Sphere. Arrows on the cuboid show surface normal directions, arrows are not included on the sphere to avoid clutter.}
    \label{fig:fept_l10_surface_energy_comparison_grains}
\end{figure}

As the table shows, the \{111\} surfaces have the lowest energy, informing our choice for three of the grain shapes, which can be seen in Figure \ref{fig:fept_l10_surface_energy_informed_grains}. As the Figure shows, these shapes are defined so that they feature predominantly \{111\} surfaces, which our surface calculations predict would make them the lowest energy structures. As a comparison, we included two extra shapes, which can be seen in Figure \ref{fig:fept_l10_surface_energy_comparison_grains}. We chose a spherical and cuboid grain for their surface area to volume ratios. Selecting shapes at the extreme ends of the surface area to volume scale, allowed us to investigate what role surface area to volume plays in energetic preferability.

From Figure \ref{fig:fept_l10_surface_energy_informed_grains}, the grain shapes are named a) Octahedron, b) Truncated-Octahedron-Major, and c) Truncated-Octahedron-Minor; from Figure \ref{fig:fept_l10_surface_energy_comparison_grains}, the grains are named a) Cuboid, b) Sphere.

\begin{figure}[!t]
    \captionsetup[subfigure]{font=Large,justification=justified,singlelinecheck=false,labelfont={rm, bf},labelformat=simple,}
    \newcommand{\vspaced}{\vspace{-0.8em}}
    \newcommand{\vvspaced}{\vspace{-1.6em}}
    \centering
    \begin{subfigure}[t]{0.4\textwidth}
        \centering
        \subcaption{}
        \vspaced
        \includegraphics[width=\textwidth]{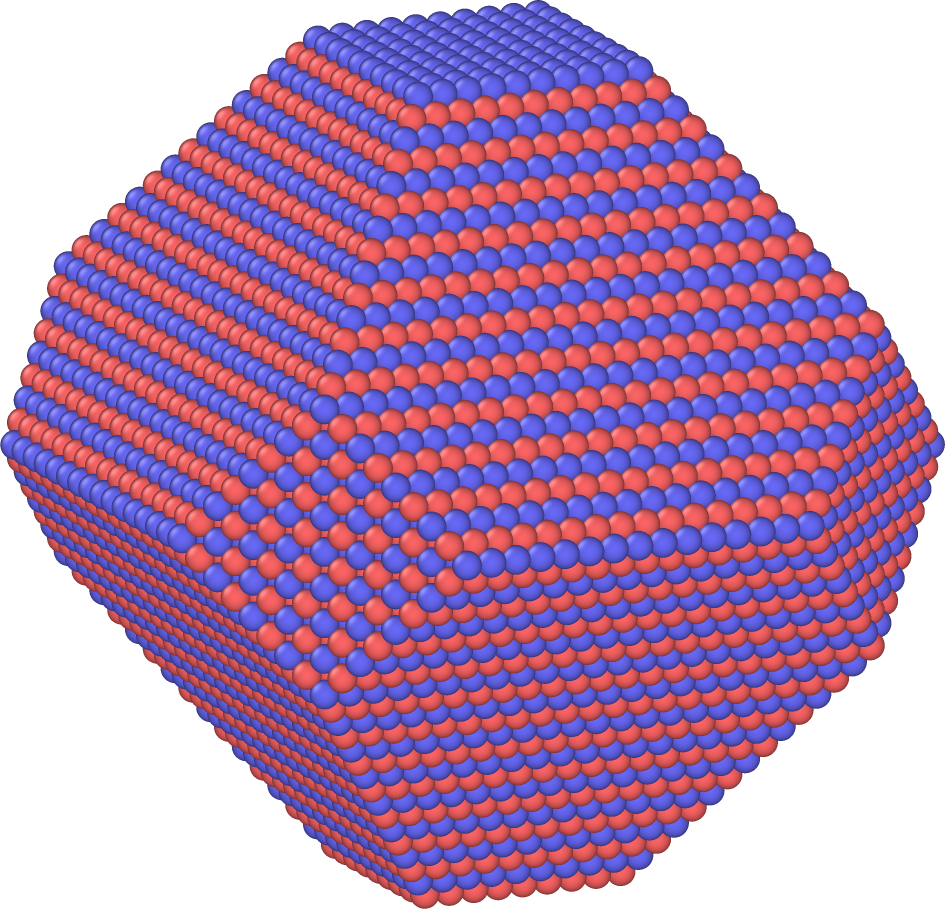}
        \label{fig:fept_l10_truncated_octahedron_minor_20000_atoms_as_built}
        \vvspaced
    \end{subfigure}
    \hspace{0.1\textwidth}
    \begin{subfigure}[t]{0.4\textwidth}
        \centering
        \subcaption{}
        \vspaced
        \includegraphics[width=\textwidth]{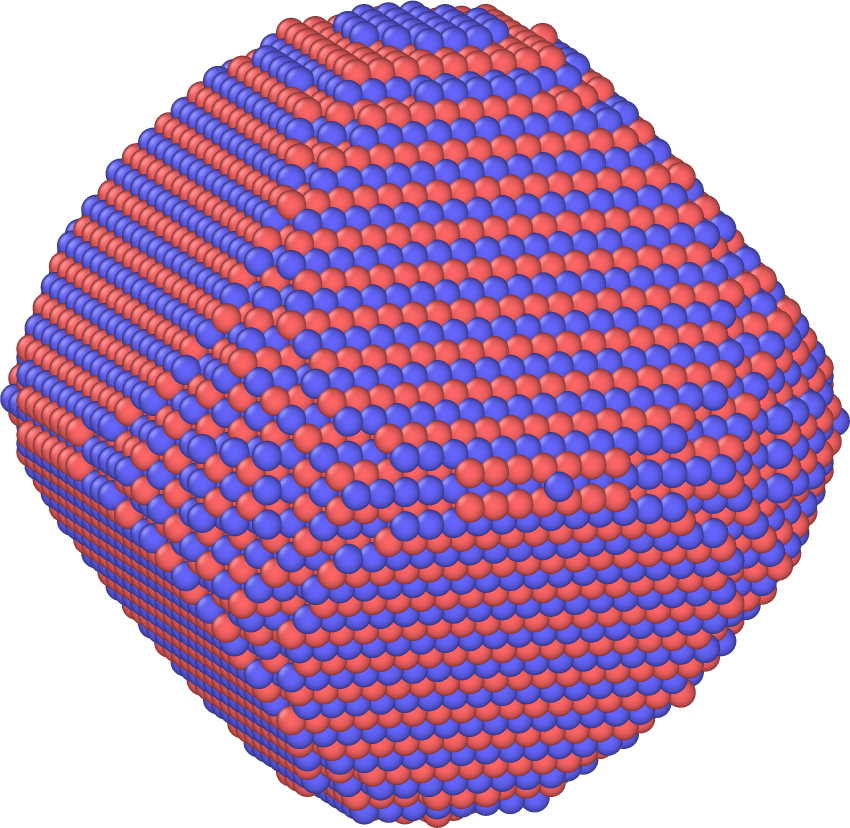}
        \label{fig:fept_l10_truncated_octahedron_minor_20000_atoms_compositionally_identical}
    \end{subfigure}
    \caption{Comparison of a 20,000 atom Truncated Octahedron Minor in its a) As built, and b) Compositionally matched, states.}
    \label{fig:fept_l10_as_built_and_compositionally_matched_grain}
\end{figure}

To compare the preferability of these shapes as a function of size, we created grains of each type ranging from 1,000 to 15,000 atoms, in steps of 1,000 atoms - equating to diameters of 3-9nm. To ensure our grains had identical compositions, we scaled the grain structures above their intended atom target, then deleted atoms from their surface to match an average composition with the desired atom number - for full details see the methodology section. An example of how this affects the structure can be seen in Figure \ref{fig:fept_l10_as_built_and_compositionally_matched_grain}, which shows how the surface of the grain structure changes to reach a matching composition. We refer to this process as composition matching, and the grains in this state as "as-matched".

\begin{figure}[!t]
    \centering
    \includegraphics[width=\textwidth]{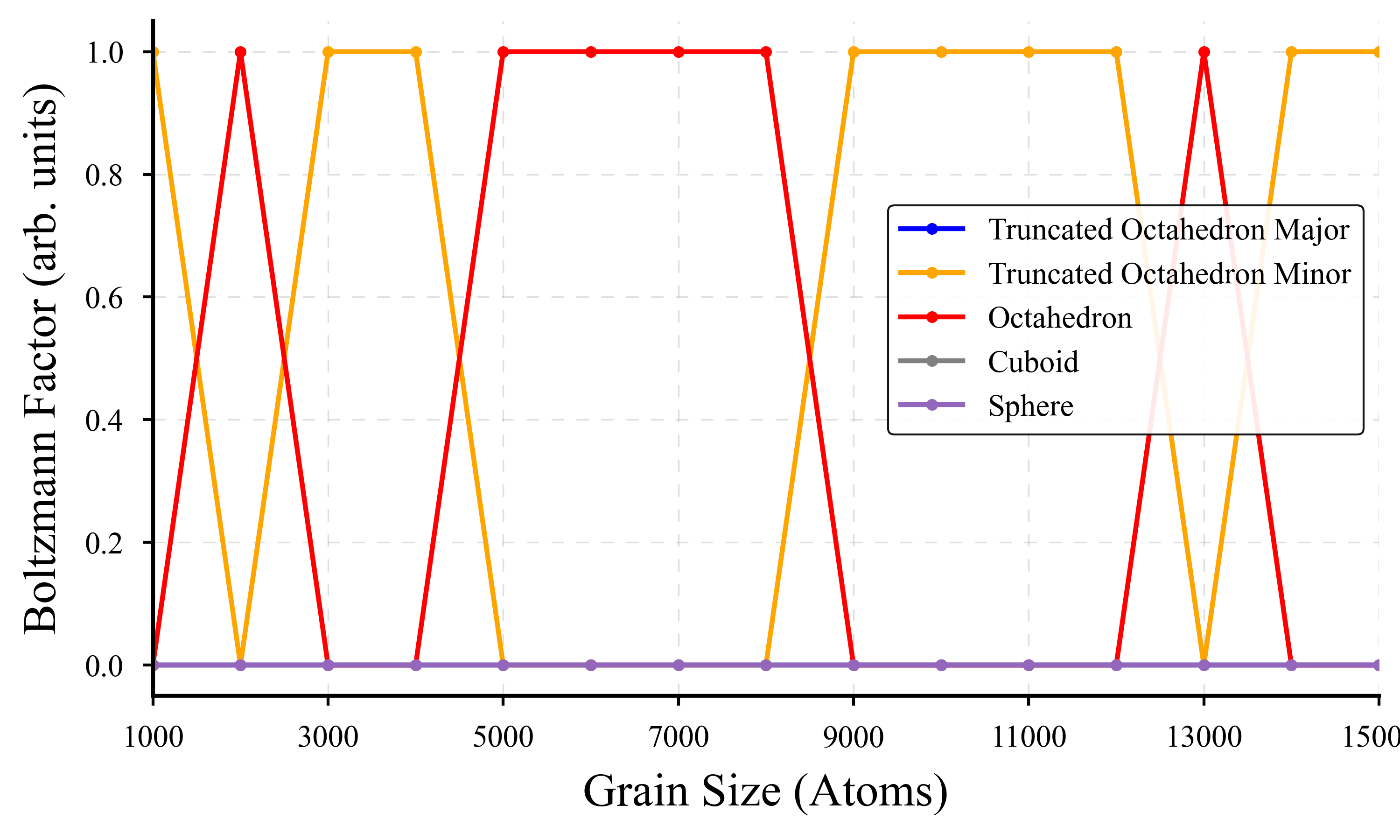}
    \caption{Boltzmann Factors for each structure over a range of 1,000 - 15,000 atoms.}
    \label{fig:fept_l10_grain_boltzmann_factor_comparison}
\end{figure}

We calculated the energy of our as-matched grains using conjugate gradient energy minimisation at 0K\cite{plimpton1995fast}. As mentioned, the grains have identical compositions, making them re-arrangements of the same underlying system. This allowed us to calculate their relative probabilities in the Boltzmann distribution by calculating their Boltzmann factors from their minimised energy. The ratio of their Boltzmann factors gave us their comparative probability with temperature - see methodology section for full details.

In Figure \ref{fig:fept_l10_grain_boltzmann_factor_comparison}, we show that, across the investigated size range, the most probable shapes are alternately the Octahedron, and the Truncated-Octahedron-Minor. This demonstrates the competition between beneficial surface selection and surface area to volume ratio, as their similar energies are due to the interplay of these two factors.

In the case of the Octahedron, on average, the surface is 87\% \{111\} planes, giving it the lowest surface energy per unit area of the all grain shapes. However, as a result of maximising the \{111\} surface, this grain shape has the worst surface area to volume ratio, giving it substantially more surface for the same volume. In contrast, the Truncated-Octahedron-Minor has an average of 71\% \{111\} surface planes, with the best surface area to volume ratio out of all the shapes - owing to the truncation of the \{111\} plane corners. The fact these two structures have similar energy, demonstrates two things, firstly, that, at these sizes, the dominant factor is surface type; secondly, that when two grains have similar percentages of the same surface type, the deciding factor is surface area to volume ratio.

For the remaining shapes, the order of preference is Sphere/Trunacted-Octahedron-Major followed by Cuboid, where Sphere and Truncated-Octahedron-Major can be considered interchangeable. In contrast to the previous pair, the Sphere and Truncated-Octahedron-Major have comparable energies due to their energetically similar surface types only. For Truncated-Octahedron-Major, this means a surface composed of, on average, 46\% \{111\} surface planes, with \{100\} and \{001\} surface planes making up the remaining surface area; for the Sphere, this means a surface composed of, on average, 53\% \{111\} planes, with a wide spread of surface planes making up the remaining surface area. The spread of the Sphere's remaining planes includes multiple high index, high energy planes that give the Sphere an overall surface energy per unit area of similar value to the Truncated-Octahedron-Major's, despite it's 7\% advantage in \{111\} surface plane coverage. The fact these two structures are alike, in both surface area and volume, further supports our conclusion that, at these length scales, surface type is the dominant factor controlling grain shape.

As expected, the Cuboid is significantly less likely than every other shape, possessing no \{111\} surface planes, except those incidentally caused by surface deletion, and a poor surface area to volume ratio.

\begin{figure}[t]
    \centering
    \includegraphics[width=0.95\textwidth]{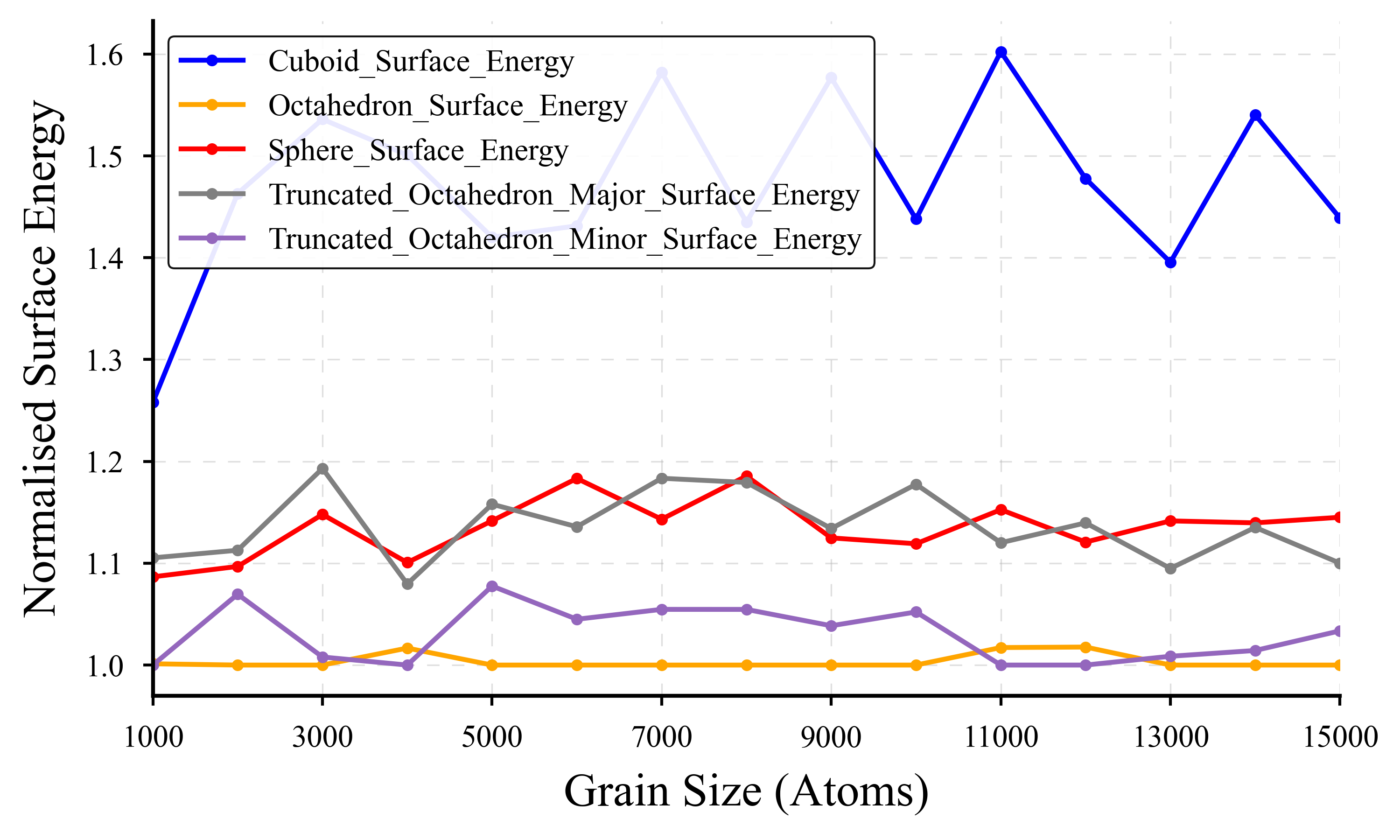}
    \caption{Normalised surface energies of the investigated grains across the range 1,000-15,000 atoms. The normalisation is different at each atom size, with the normalising value being the grain with the lowest surface energy at each point.}
    \label{fig:fept_l10_normalised_surface_energies}
\end{figure}

To demonstrate the predominant role of surface type, we calculated the surface energy for each of the grains across the size range. Figure \ref{fig:fept_l10_normalised_surface_energies} shows these energies, normalised by the minimum surface energy at every size. As can be seen, the preference order we outlined above is reflected in the surface energies. For grains of identical composition, surface energy is the singular defining factor and for grains of this size range, surface type is the dominant factor affecting surface energy.

\section*{Discussion}

We have developed a model which uses crystal symmetries to predict grain growth and shape as a function of size - measured in total atoms. The methodology we have developed is applicable to any crystal structure, and we anticipate using it in the future to study the probable shapes of Rare-Earth Intermetallics, particularly \ch{RT12} phases, whose stability, and therefore manufacturability, is dependent on the microstructure surrounding its grains.

Magnetron-sputtering-based gas-phase condensation experiments, by Qui and Wang\cite{qiu2007tuning}, show FePt L1$_0$ has a strong preference for the Octahedron, and Truncated-Octahedron-Minor shapes, verifying our model and methodology.

Further to this, the surface energy basis of our model allows us to make predictions about the microstructure of as-annealed FePt L1$_0$ thin films, which we predict will organise into FePt L1$_0$ configurations that maximise the \{111\} plane. As the underlayer that supports the FePt L1$_0$ thin film, pre-disposes the crystal structure to align its \{001\} direction perpendicular to the plane, for isolated FePt the reordering will cause the as-deposited material to conglomerate into hill like structures, whose slopes are composed of \{111\} surface planes. A review of the literature shows that this does occur\cite{wicht2016atomic}, which reinforces the validity of our model and opens the possibility that experimentalists and modellers will be able to produce quantitative data on the physical reasoning behind the texturing process.

With development, the model will evolve from a single methodology into a suite of tools, which we are building into a Python package: \href{https://github.com/Connor56/Grain_Modeller}{Grain Modeller}. We believe our package has two benefits, firstly, theoreticians and modellers can use it to calculate realistic grain shapes and reduce assumptions in their work; secondly, experimentalists can use it to create models that test grain shape preferability under a range of conditions, allowing them to fine tune their apparatus for a greater degree of structural control, before undertaking an experiment.

We believe that this extension to the material science toolbox will increase our communities collective ability to control microstructure and increase the speed of magnetic material development.

\section*{Methods}

\subsection*{Grain Creation}

To create FePt L1$_\text{0}$ grains, we produced a Python package, for defining grain shapes, creating grain shapes, matching grain compositions, and creating LAMMPS data and input files.

\paragraph*{Grain Shape Definition} \

To make grain shapes reproducible over a wide range of sizes, we developed a method of defining them abstractly, using a unitcell, a set of cuts, and a repeat ratio. For clarity, these are defined below:

\begin{itemize}
    \item Unitcells, are the basis of the grain, defined by three lattice vectors $\bvec{a}$, $\bvec{b}$, $\bvec{c}$ and a list of constituent atoms in fractional coordinates.
    \item Cuts, are defined by a plane normal and a plane point, both given in fractional coordinates. For example, the cut with normal [111] and plane point (0, 0, 1) would cut along the (111) plane that passes through the point at the grains maximum extent in the $\bvec{c}$ lattice direction.
    \item Repeat Ratio defines the ratio between unitcell repeats in the three lattice directions. For example a repeat ratio of [1, 2, 1] would repeat the unitcell two times in the $\bvec{b}$ direction for every single repeat in the $\bvec{a}$ and $\bvec{c}$ directions. 
\end{itemize}

Together, Cuts and Repeat Ratio define a scalable general shape. To create grains of definite sizes, we used an integer scaling factor, which works in the two following ways:
\begin{enumerate}
    \item Multiplies the repeat ratio to get the unitcell repeats in each direction. For example if the repeat ratio is [1, 2, 1] and you have a scaling factor of 10, the final repetitions in each direction are [10, 20, 10].
    \item Scales the plane point of each cut. For example, if a cut is defined by the plane normal [111] and the plane point (0, 0, 1), having a final number of unitcell repeats equal to: [10, 20, 10], scales the plane point to (0, 0, 10).
\end{enumerate}

With size factor, we have all the required information to build a grain. We create the grain by using the repeat ratio to turn the unitcell into a supercell, and cutting the supercell along the defined planes.

\paragraph{Composition Matching} \

To compare grain preference using Boltzmann factors they must have the exact same chemical composition.

To achieve this, we used a composition matching algorithm. The process starts with defining two or more grain shapes, then selecting a grain size target, measured in atoms. Each grain shape is scaled so that it's close to, but above, the target number of atoms. We call grains in this state "as built".

Each as built grain's element ratio is used to define a best fit composition. Using FePt L1$_0$ as an example, grain ratios of 49:51, 49.5:50.5, and 51:49 (Fe:Pt), give you an average of 49.83:50.17, which is used to decide the number of each atom type in the grains. Continuing the example, for grains with 10,000 atoms total, the best fit composition would be 0.4983$\times$10,000 = 4983 iron atoms, and 0.5017$\times$10,000 = 5017 platinum atoms.

To match the compositions, atoms must be removed from the surface of the grains; defining a best fit composition ensures that each grain's surface will always have the desired atoms for removal.

Once the best fit composition has been defined, atoms are progressively removed from the surface of the as built grains, until they match the desired composition. Atoms are preferentially removed from the least energetically favourable positions first, starting with corner atoms, then edge atoms, and finally general surface atoms. This produces physically realistic grains.

An example of the as built, and the compositionally matched final grain can be seen in Figure \ref{fig:fept_l10_as_built_and_compositionally_matched_grain}.

As the figure shows, the as built grain has clean surfaces with defined edges, in contrast the compositionally matched grain has rounded edges, although it maintains the same underlying morphology.

\paragraph{FePt Grain Shapes} \

\begin{figure}[!t]
    \captionsetup[subfigure]{font=Large,justification=justified,singlelinecheck=false,labelfont={rm, bf},labelformat=simple,}
    \newcommand{\vspaced}{\vspace{-0.8em}}
    \newcommand{\vvspaced}{\vspace{-1.6em}}
    \centering
    \begin{subfigure}[t]{0.48\textwidth}
        \centering
        \subcaption{}
        \vspaced
        \includegraphics[width=\textwidth]{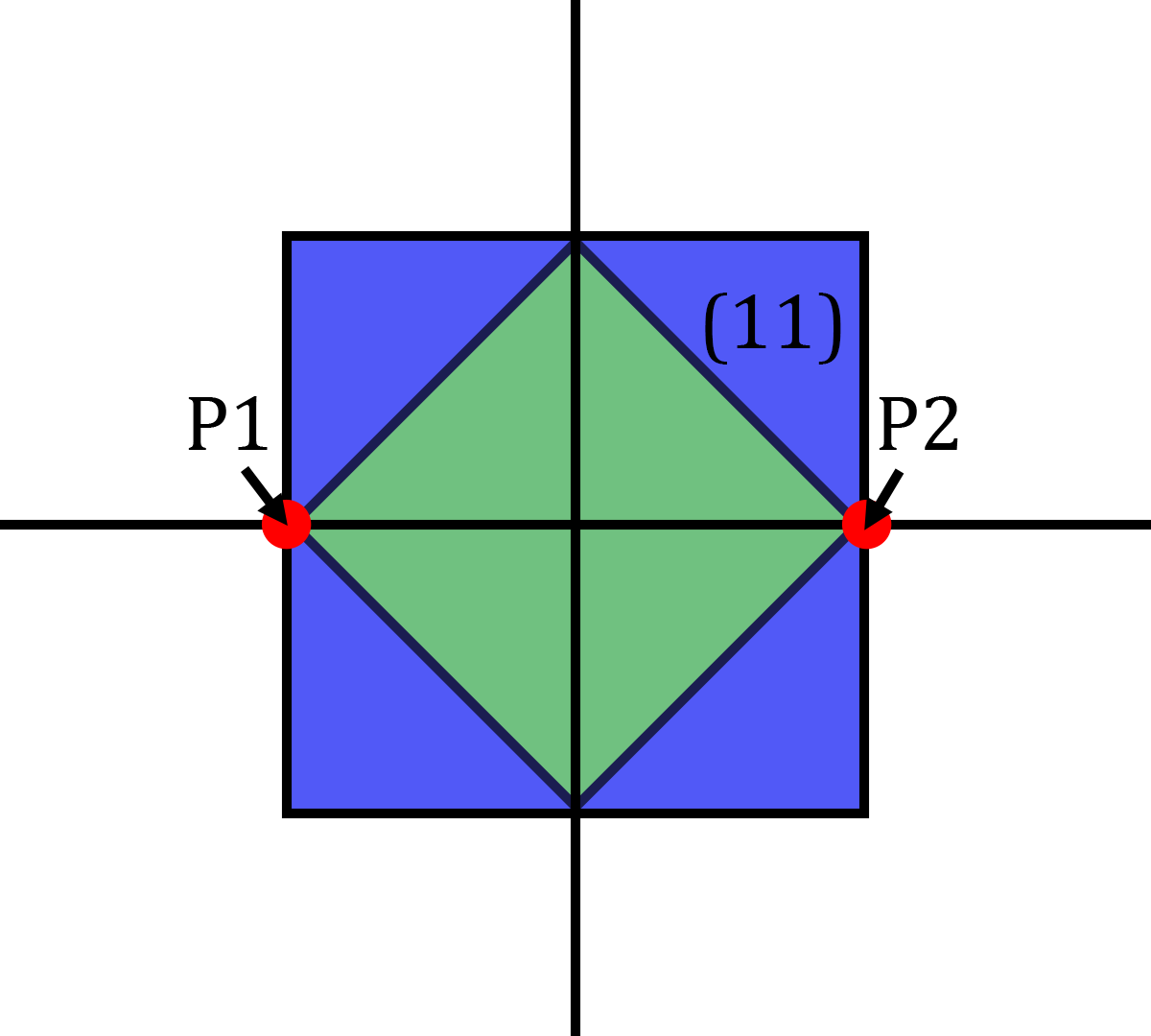}
        \label{fig:fept_l10_octahedron_plane_cuts_from_point_placement_2d}
        \vvspaced
    \end{subfigure}
    \hspace{0.02\textwidth}
    \begin{subfigure}[t]{0.48\textwidth}
        \centering
        \subcaption{}
        \vspaced
        \includegraphics[width=\textwidth]{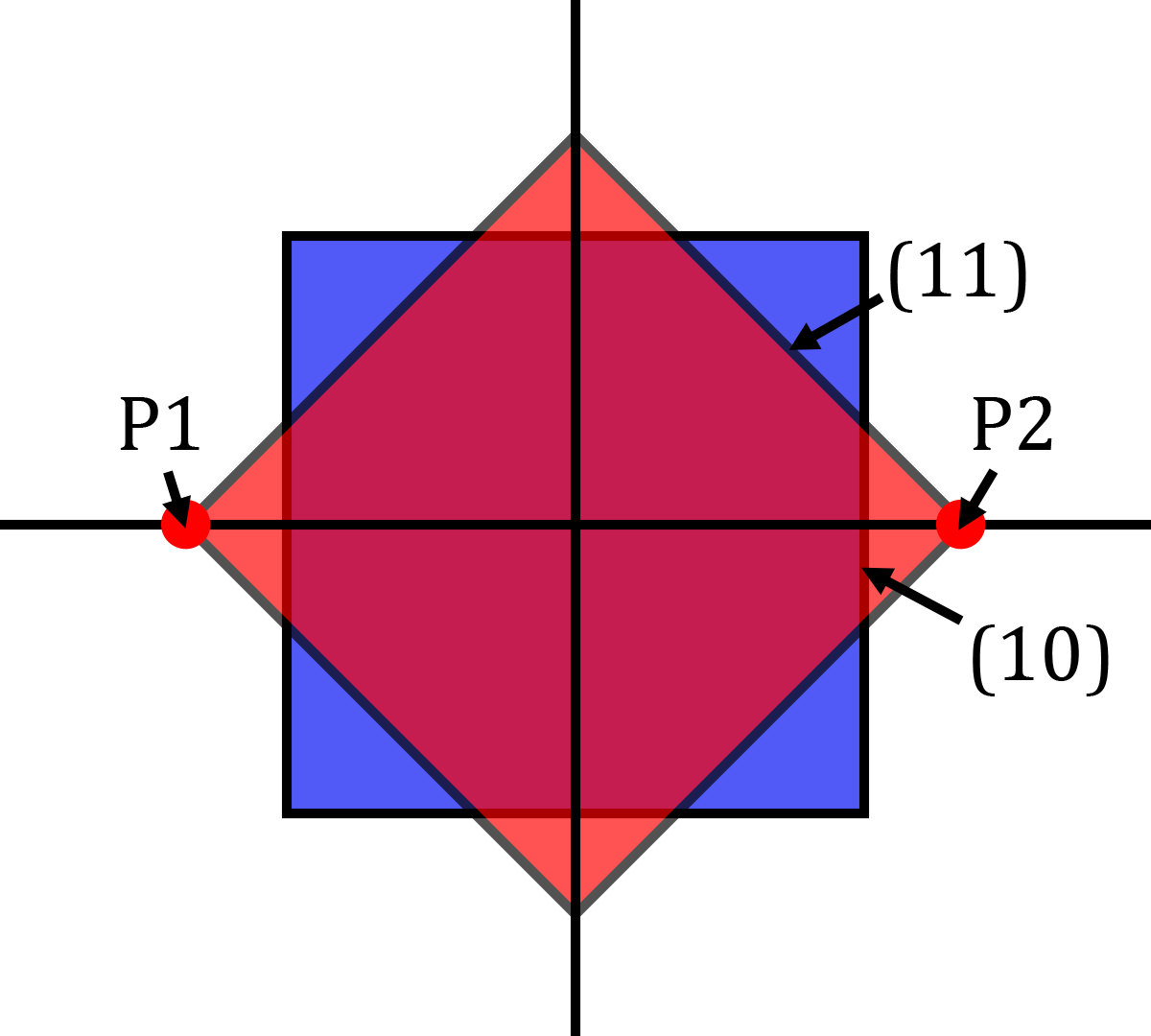}
        \label{fig:fept_l10_truncated_octahedron_minor_plane_cuts_from_point_placement_2d}
    \end{subfigure}
    \begin{subfigure}[t]{0.48\textwidth}
        \centering
        \subcaption{}
        \vspaced
        \includegraphics[width=\textwidth]{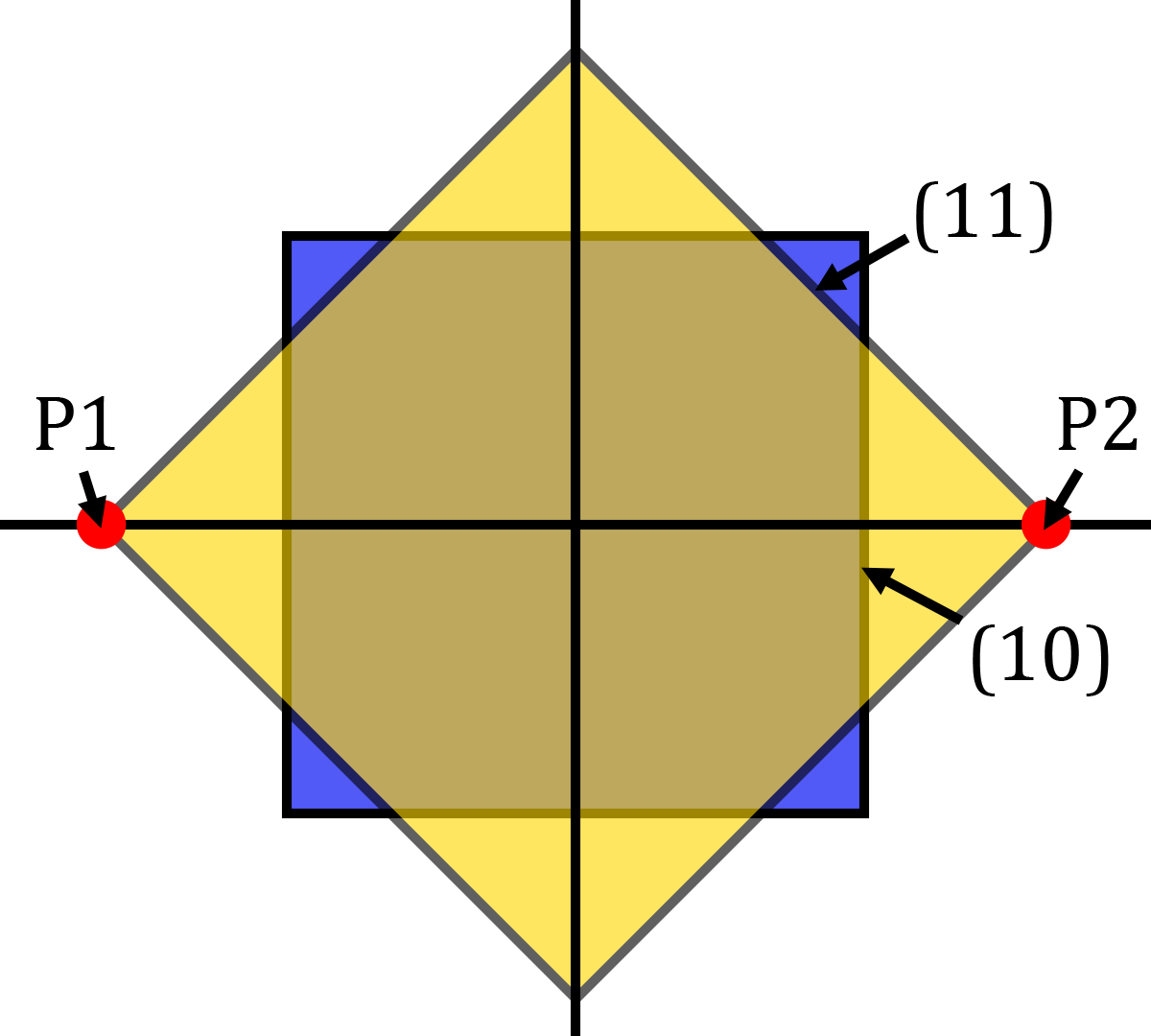}
        \label{fig:fept_l10_truncated_octahedron_major_plane_cuts_from_point_placement_2d}
    \end{subfigure}
    \caption{2D example of how point placement affects grain morphology, when cuts are made along the \{11\} plane set. The blue cube represents the original cubic supercell and the coloured diamonds represent the area left untouched by the \{11\} cuts. The final grain shape is given by the area of the blue cube that is covered by the coloured diamond. The points which define the placement of cuts are noted in each diagram. The figure shows: a) The 2D representation of the Octahedron shape, b) The 2D representation of the Truncated Octahedron Minor shape, and c) The 2D representation of the Truncated Octahedron Major shape.}
    \label{fig:fept_l10_cut_plane_point_placement_2D}
\end{figure}

To choose our grain morphologies, we calculated the surface energies of L1$_0$'s low index planes to see which should dominate our grains' surfaces. Our calculated surface energies can be see in Table \ref{tab:fept_l10_surface_energies} alongside surface calculations taken from a paper by Kim et al.\cite{kim2006origin}.

Our calculations showed that \{111\} surfaces have the lowest energy, due to a greater number of interplanar bonds (6 vs 4).

Using this as a basis, all our grain shapes used cuts along the \{111\} planes to expose as much of this preferential surface as possible; \{100\} and \{001\} planes, which naturally accompany \{111\} plane cuts, were left to compose the remaining surface. As noted in the results, we looked at five types of grain shape, three with \{111\} dominated surfaces: Octahedron, Truncated Octahedron Minor, and Truncated Octahedron Major; and two with energetically unpreferable surfaces: Sphere and Cuboid. The five grain types can be seen in Figures \ref{fig:fept_l10_surface_energy_informed_grains} and \ref{fig:fept_l10_surface_energy_comparison_grains}.

For each grain, we used an FePt L1$_0$ unitcell with the following parameters: $a = b = 3.83$\AA{}, $c = 3.711$\AA{}, and $\alpha = \beta = \gamma = 90\degree$\cite{thiele1998perpendicular}, where $a$, $b$, $c$ are the lattice parameters, and $\alpha$, $\beta$, $\gamma$ are the angles between them. Each set of \{111\} plane cuts, used to create the three none extreme grain shapes, used two plane points, $P_1$ and $P_2$, selected along the positive and negative $a$ direction. The points were chosen to obey the property: $|\bvec{P_1}| = |\bvec{P_2}|$, assuring the as built grains had a symmetrical structure.

The shape of each grain was decided by changing the distance of each point from the centre of the supercell. Figure \ref{fig:fept_l10_cut_plane_point_placement_2D} demonstrates how point placement affects the final shape of 2D as built grains.

As can be seen in Figure \ref{fig:fept_l10_cut_plane_point_placement_2D}, the area of the cube covered by the diamond creates a similar grain shape to our 3D cases projected onto a 2D plane. The edges of the diamond passing through the plane points can be thought of as the \{11\} planes associated with that particular point. In 2D there are only two planes associated with each point, but for the 3D case each point has four \{111\} planes associated with it.

\subsection*{Simulation Methodology}

We used MEAM potentials within the LAMMPS molecular dynamics package to perform energy minimisation calculations.

\paragraph{MEAM Potentials} \

The constants for our MEAM potentials are given in Table's \ref{tab:meam_potential_constants_fe_pt} and \ref{tab:meam_potential_constants_shared}.
\begin{table}[!b]
    \centering
    \begin{tabular}{|c|c|c|}
        \hline
        Parameter & Fe & Pt \\
        \hline
        E$_c$ & 4.29 & 5.77 \\
        R$_e$ & 2.867 & 3.92 \\
        B & 1.08 & 1.80 \\
        A & 0.56 & 0.90 \\
        $\beta^{(0)}$ & 4.15 & 4.92 \\
        $\beta^{(1)}$ & 1.0 & 2.2 \\
        $\beta^{(2)}$ & 1.0 & 6.0 \\
        $\beta^{(3)}$ & 1.0 & 2.2 \\
        $t^{(0)}$ & 1.0 & 1.0 \\
        $t^{(1)}$ & 2.6 & 3.94 \\
        $t^{(2)}$ & 1.8 & -2.20 \\
        $t^{(3)}$ & -7.2 & 3.84 \\
        \hline
    \end{tabular}
    \caption{MEAM potential parameters for iron and platinum. E$_c$ is the reference structure cohesive energy, R$_e$ is the reference structure equilibrium bond distance, B is the Bulk modulus, A is a model parameter that scales the screening functions, $\beta^{(l)}$ $l$=1-3 are scaling parameters controlling the form of the original EAM partial electron density functions: $\rho_i^{a(l)}$, and $t^{(0)}$ which scales the contribution of each of the MEAM partial electron densities: $\rho_i^{(l)}$ $l$=1-3.}
    \label{tab:meam_potential_constants_fe_pt}
    \vspace{2em}
\end{table}
\begin{table}[!t]
    \def\arraystretch{1.1}
    \centering
    \begin{tabular}{|c|c|}
        \hline
        Parameter & Value \\
        \hline
        \ch{C_{min}}(Fe-Fe-Fe) &  0.36 \\
        \ch{C_{max}}(Fe-Fe-Fe) &  2.80 \\
        \ch{C_{min}}(Pt-Pt-Pt) &  1.53 \\
        \ch{C_{max}}(Pt-Pt-Pt) &  2.80 \\
        \ch{C_{min}}(Pt-Fe-Fe) &  0.36 \\
        \ch{C_{max}}(Pt-Fe-Fe) &  2.80 \\
        \ch{C_{min}}(Fe-Pt-Pt) &  1.53 \\
        \ch{C_{max}}(Fe-Pt-Pt) &  2.80 \\
        \ch{C_{min}}(Fe-Fe-Pt) &  0.844 \\
        \ch{C_{max}}(Fe-Fe-Pt) &  2.80 \\
        \ch{C_{min}}(Pt-Fe-Pt) &  0.844 \\
        \ch{C_{max}}(Pt-Fe-Pt) &  2.80 \\
        \ch{E_{c}}(Fe, Pt) & 5.86 \\
        \ch{R_{e}}(Fe, Pt) & 2.781 \\
        \hline
    \end{tabular}
    \caption{The MEAM potential constants governing interactions between iron and platinum. The \ch{C_{min}} and \ch{C_{max}} constants are the maximum and minimum values for the screening functions when atoms are screened by an intervening atom, for example \ch{C_{min}}(Pt-Fe-Pt) is the maximum screening function due to an iron atom screening the interaction of two platinum interactions. \ch{E_{c}}(Fe, Pt) is the energy of the iron - platinum reference structure, and \ch{R_{e}}(Fe, Pt) is the equilibrium bond distance of the iron - platinum reference structure.}
    \label{tab:meam_potential_constants_shared}
\end{table}
The potentials were taken from a paper by Kim et al.\cite{kim2006origin} and Table \ref{tab:fept_l10_surface_energies} shows a comparison between our calculated surface energies and Kim et al.'s surface energies. The comparison shows that our calculations are in good agreement with Kim et al. in the \{111\}, \{101\} and \{110\} planes, but in poor agreement for the \{100\} and \{001\} planes. The exact reason for this isn't clear, although it's possible the discrepancy is due to different surface energy calculation methods. However, as our work is primarily intended to demonstrate a methodology and the trend of the potentials is roughly the same, we deemed the potentials accurate enough for this work.

\paragraph{Simulation Methodology} \

Before performing energy minimisation, each grain was equilibrated in an NPT ensemble to avoid convergence to a local minimum. The NPT ensemble is set at 1K with a pressure of 0Pa and evolved the system over 15,000 time steps. The simulation's time steps were 0.001ps, making this phase 15ps long. Minimisation used the standard conjugate gradient algorithm in LAMMPS and occured over a minimum of 10,000 time steps to ensure the maximum level of covergence. Generally, convergence was reached much sooner.

\begin{figure}[!h]
    \centering
    \includegraphics[width=\textwidth]{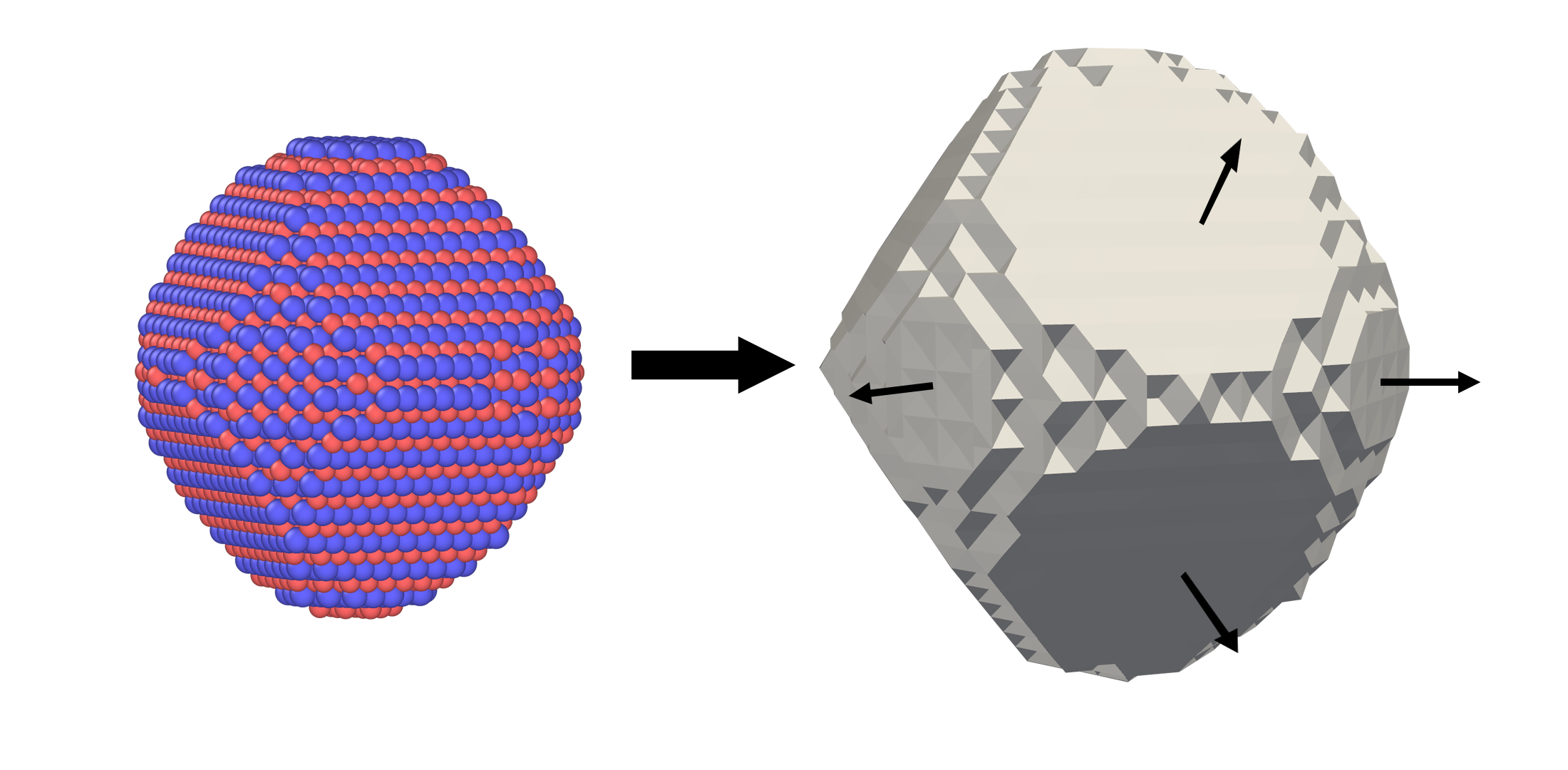}
    \caption{Left) Atomistic visualisation of an FePt L10 Truncated Octahedron Minor grain, Right) Surface of the grain extracted by PyVista, arrows out of the surfaces indicating the plane normals.}
    \label{fig:fept_l10_surface_extraction}
\end{figure}

\subsection*{Probability Calculations}

To calculate the probability of each grain shape we used their minimised energy to produce their Boltzmann factors via the following equation:

\begin{equation}
    p_i \propto e^{\frac{-\epsilon_i}{k_BT}}
\end{equation}

where $p_i$ is the probability of grain $i$, $\epsilon_i$ is the minimised energy of grain $i$, $k_B$ is the Boltzmann constant, and $T$ is the temperature. Normalising each grain's Boltzmann factor by the minimum energy grain's Boltzmann factor gave us their relative probabilities. Plotting this gave us Figure \ref{fig:fept_l10_grain_boltzmann_factor_comparison}.

As they are compositionally identical, the grains are rearrangements of the same system. Therefore, taking the ratio of their Boltzmann factors is the same as taking the ratio of their probabilities, and this method allowed us to make predictions about the most probable structure.

\subsection*{Surface Energy Calculations}

To investigate what role surface energy plays in shape probability probability, we developed an algorithm that extracts the surface from the grain shapes.

We used PyVista\cite{sullivan2019pyvista}, which is a wrapper around VTK\cite{schroeder1998visualization}, to extract the surfaces using Delaunay triangulation\cite{lee1980two}. Used on each grain, this algorithm gave us a collection of triangles (2D simplexes) that made up the grain's surface. We used the triangle's points to calculate their plane normals and area. Grouping the triangles by their plane normals allowed us to calculate the surface area associated with each plane, e.g. \{111\}, \{001\}, etc.

Finally, to get the each grain's surface energy, we multiplied their surface areas by our calculated surface energies per unit area shown in Table \ref{tab:fept_l10_surface_energies}. This process is shown visually in Figure \ref{fig:fept_l10_surface_extraction} and the code can be found in the modules contained in this \href{https://github.com/Connor56/Grain_Modeller/tree/main/grain_modeller/analyse/grains}{folder}, in the Grain Modeller Python package.

\bibliography{references}

\section*{Author contributions statement}

G.H. directed the project, C.S. developed the Python Package, C.S. wrote the Python package, C.S. and G.H. conceived the models, C.S. built the models, C.S. ran the simulations, C.S. and G.H. analysed the results, and C.S. and G.H. wrote the paper. All authors reviewed the manuscript.

\section*{Acknowledgements}

Thank you to Toyota for sponsoring this work.

\section*{Additional information}

The authors declare no competing interests.

\end{document}